# Temperature-dependent Gilbert damping of $Co_2FeAl$ thin films with different degree of atomic order

Ankit Kumar[1]*, Fan Pan[2,3], Sajid Husain[4], Serkan Akansel[1], Rimantas Brucas[1], Lars Bergqvist[2,3], Sujeet Chaudhary[4], and Peter Svedlindh[1#]

[1]Department of Engineering Sciences, Uppsala University, Box 534, SE-751 21 Uppsala, Sweden

[2]Department of Applied Physics, School of Engineering Sciences, KTH Royal Institute of Technology, Electrum 229, SE-16440 Kista, Sweden

[3]Swedish e-Science Research Center, KTH Royal Institute of Technology, SE-10044 Stockholm, Sweden

[4]Department of Physics, Indian Institute of Technology Delhi, New Delhi-110016, India

**ABSTRACT**

Half-metallicity and low magnetic damping are perpetually sought for in spintronics materials and full Heusler alloys in this respect provide outstanding properties. However, it is challenging to obtain the well-ordered half-metallic phase in as-deposited full Heusler alloys thin films and theory has struggled to establish a fundamentals understanding of the temperature dependent Gilbert damping in these systems. Here we present a study of the temperature dependent Gilbert damping of differently ordered as-deposited $Co_2FeAl$ full Heusler alloy thin films. The sum of inter- and intraband electron scattering in conjunction with the finite electron lifetime in Bloch states govern the Gilbert damping for the well-ordered phase in contrast to the damping of partially-ordered and disordered phases which is governed by interband electronic scattering alone. These results, especially the ultralow room temperature intrinsic damping observed for the well-ordered phase provide new fundamental insights to the physical origin of the Gilbert damping in full Heusler alloy thin films.

# INTRODUCTION

The Co-based full Heusler alloys have gained massive attention over the last decade due to their high Curie temperature and half-metallicity; 100% spin polarization of the density of states at the Fermi level [1-2]. The room temperature half-metallicity and low Gilbert damping make them ideal candidates for magnetoresistive and thermoelectric spintronic devices [3]. $Co_2FeAl$ (CFA), which is one of the most studied Co-based Heusler alloys, belongs to the $Fm\overline{3}m$ space group, exhibits half-metallicity and a high Curie temperature (1000 K) [2, 4]. In CFA, half-metallicity is the result of hybridization between the $d$ orbitals of Co and Fe. The $d$ orbitals of Co hybridize resulting in bonding ($2e_g$ and $3t_{2g}$) and non-bonding hybrids ($2e_u$ and $3t_{1u}$). The bonding hybrids of Co further hybridise with the $d$ orbitals of Fe yielding bonding and anti-bonding hybrids. However, the non-bonding hybrids of Co cannot hybridise with the $d$ orbitals of Fe. The half-metallic gap arises from the separation of non-bonding states, i.e. the conduction band of $e_u$ hybrids and the valence band of $t_{1u}$ hybrids [5, 6]. However, chemical or atomic disorder modifies the band hybridization and results in a reduced half-metallicity in CFA. The ordered phase of CFA is the $L2_1$ phase, which is half-metallic [7]. The partially ordered B2 phase forms when the Fe and Al atoms randomly share their sites, while the disordered phase forms when Co, Fe, and Al atoms randomly share all the sites [5-8]. These chemical disorders strongly influence the physical properties and result in additional states at the Fermi level therefore reducing the half-metallicity or spin polarization [7, 8]. It is challenging to obtain the ordered $L2_1$ phase of Heusler alloys in as-deposited films, which is expected to possess the lowest Gilbert damping as compared to the other phases [4, 9-11]. Therefore, in the last decade several attempts have been made to grow the ordered phase of CFA thin films employing different methods [4, 9-13]. The most successful attempts used post-deposition annealing to reduce the anti-site disorder by a thermal activation process [4]. The observed value of the Gilbert damping for ordered thin films was found to lie in the range of 0.001-0.004 [7-13]. However, the requirement of post-deposition annealing might not be compatible with the process constraints of spintronics and CMOS devices. The annealing treatment requirement for the formation of the ordered phase can be circumvented by employing energy enhanced growth mechanisms such as ion beam sputtering where the sputtered species carry substantially larger energy, ~20 eV, compared to other deposition techniques [14, 15]. This higher energy of the sputtered species enhances the ad-atom mobility during coalescence of nuclei in the initial stage of the thin film growth, therefore enabling the formation of the ordered phase. Recently we have reported growth of the ordered CFA phase on potentially advantageous Si substrate using ion beam sputtering. The samples deposited in the range of 300°C to 500°C substrate temperature exhibited nearly equivalent I(002)/I(004) Bragg diffraction intensity peak ratio, which confirms at least B2 ordered phase as it is difficult to identify the formation of the L2$_1$ phase only by X-ray diffraction analysis [16].

Different theoretical approaches have been employed to calculate the Gilbert damping in Co-based full Heusler alloys, including first principle calculations on the basis of (i) the torque correlation model [17], (ii) the fully relativistic Korringa-Kohn-Rostoker model in conjunction with the coherent potential approximation and the linear response formalism [8], and (iii) an approach considering different exchange correlation effects using both the local spin density approximation including the Hubbard U and the local spin density approximation plus the dynamical mean field theory approximation [7]. However, very little is known about the temperature dependence of the Gilbert damping in differently ordered Co-based Heusler alloys and a unifying consensus between theoretical and experimental results is still lacking. In this study we report the growth of differently ordered phases, varying from disordered to well-ordered phases, of as-deposited CFA thin films grown on Si employing ion beam sputtering and subsequently the detailed temperature dependent measurements of the Gilbert damping. The observed increase in intrinsic Gilbert damping with decreasing temperature in the well-ordered sample is in contrast to the continuous decrease in intrinsic Gilbert damping with decreasing temperature observed for partially ordered and disordered phases. These results are satisfactorily explained by employing spin polarized relativistic Korringa-Kohn-Rostoker band structure calculations in combination with the local spin density approximation.

**SAMPLES & METHODS**

Thin films of CFA were deposited on Si substrates at various growth temperatures using ion beam sputtering system operating at 75W RF ion-source power ($P_{ion}$). Details of the deposition process as well as structural and magnetic properties of the films have been reported elsewhere [16]. In the present work to study the temperature dependent Gilbert damping of differently ordered phases (L2$_1$ and B2) we have chosen CFA thin films deposited at 573K, 673K and 773K substrate temperature ($T_S$) and the corresponding samples are named as LP573K, LP673K and LP773K, respectively. The sample thickness was kept constant at 50 nm and the samples were capped with a 4 nm thick Al layer. The capping layer protects the films by forming a 1.5 nm thin protective layer of Al$_2$O$_3$. To obtain the A2 disordered CFA phase, the thin film was deposited at 300K on Si employing 100W ion-source power, this sample is referred to as HP300K. Structural and magnetic properties of this film are presented in Ref. [18]. The absence of the (200) diffraction peak in the HP300K sample [18] reveals that this sample exhibits the A2 disordered structure. The appearance of the (200) peak in the LP series samples clearly indicates at least formation of B2 order [16]. Employing the Webster model along with the analysis approach developed by Takakura et al. [19] we have calculated the degree of B2 ordering in the samples, $S_{B2} = \sqrt{\frac{I_{200}/I_{220}}{I_{200}^{full\ order}/I_{220}^{full\ order}}}$, where $I_{200}/I_{220}$ is the experimentally obtained intensity ratio of the (200) and (220) diffractions and $I_{200}^{full\ order}/I_{220}^{full\ order}$ is the theoretically calculated intensity ratio for fully ordered B2 structure in polycrystalline films [20]. The estimated values of $S_{B2}$ for the LP573, LP673, and LP773 samples are found to be ~ 90 %, 90% and 100%, respectively, as presented in Ref. [20]. The $I_{200}/I_{400}$ ratio of the (200) and (400) diffraction peaks for all LP series samples is ~ 30 %, which compares well with the theoretical value for perfect B2 order [21, 22]. Here it is important to note that the L21 ordering parameter, $S_{L2_1}$, will take different values depending on the degree of B2 ordering. $S_{L2_1}$ can be calculated from the $I_{111}/I_{220}$ peak ratio in conjunction with the $S_{B2}$ ordering parameter [19]. However, in the recorded grazing incident XRD spectra on the polycrystalline LP samples (see Fig. 1 of Ref. [16]) we did not observe the (111) peak. This could be attributed to the fact that theoretical intensity of this peak is only around two percent of the (220) principal peak. The appearance of this peak is typically observed in textured/columnar thicker films [19, 23]. Therefore, here using the experimental results of the Gilbert damping, Curie temperature and saturation magnetization, in particular employing the temperature dependence of the Gilbert damping that is very sensitive to the amount of site disorder in CFA films, and comparing with corresponding results obtained from first principle calculations, we provide a novel method for determining the type of crystallographic ordering in full Heusler alloy thin films.

The observed values of the saturation magnetization ($\mu_0 M_S$) and coercivity ($\mu_0 H_{ci}$), taken from Refs. [16, 18] are presented in Table I. The temperature dependence of the magnetization was recorded in the high temperature region (300–1000K) using a vibrating sample magnetometer in an external magnetic field of $\mu_0 H = 20$ mT. An ELEXSYS EPR spectrometer from Bruker equipped with an X-band resonant cavity was used for angle dependent in-plane ferromagnetic resonance (FMR) measurements. For studying the temperature dependent spin dynamics in the magnetic thin films, an in-house built out-of-plane FMR setup was used. The setup, using a Quantum Design Physical Properties Measurement System covers the temperature range 4 – 350 K and the magnetic field range ±9T. The system employs an Agilent N5227A PNA network analyser covering the frequency range 1 – 67 GHz and an in-house made coplanar waveguide. The layout of the system is shown in Fig. 1. The complex transmission coefficient ($S_{21}$) was recorded as a function of magnetic field for different frequencies in the range 9-20 GHz and different temperatures in the range 50-300 K. All FMR measurements were recorded keeping constant 5 dB power.

To calculate the Gilbert damping, we have the used the torque–torque correlation model [7, 24], which includes both intra- and interband transitions. The electronic structure was obtained from the spin polarized relativistic Korringa-Kohn-Rostoker (SPR-KKR) band structure method [24, 25] and the local spin density approximation (LSDA) [26] was used for the exchange correlation potential. Relativistic effects were taken into account by solving the Dirac equation for the electronic states, and the atomic sphere approximation (ASA) was employed for the shape of potentials. The experimental bulk value of the lattice constant [27] was used. The angular momentum cut-off of $l_{max} = 4$ was used in the multiple-scattering expansion. A k-point grid consisting of ~1600 points in the irreducible Brillouin zone was employed in the self-consistent calculation while a substantially more dense grid of ~60000 points was employed for the Gilbert damping calculation. The exchange parameters $J_{ij}$ between the atomic magnetic moments were calculated using the magnetic force theorem implemented in the Liechtenstein-Katsnelson-Antropov-Gubanov (LKAG) formalism [28, 29] in order to construct a parametrized model Hamiltonian. For the B2 and L2$_1$ structures, the dominating exchange interactions were found to be between the Co and Fe atoms, while in A2 the Co-Fe and Fe-Fe interactions are of similar size. Finite temperature properties such as the temperature dependent magnetization was obtained by performing Metropolis Monte Carlo (MC) simulations [30] as implemented in the UppASD software [31, 32] using the parametrized Hamiltonian. The coherent potential approximation (CPA) [33, 34] was applied not only for the treatment of the chemical disorder of the system, but also used to include the effects of quasi-static lattice displacement and spin fluctuations in the calculation of the temperature dependent Gilbert damping [35–37] on the basis of linear response theory [38].

**RESULTS & DISCUSSION**

### A. Magnetization vs. temperature measurements

Magnetization measurements were performed with the ambition to extract values for the Curie temperature ($T_C$) of CFA films with different degree of atomic order; the results are shown in Fig. 2(a). Defining $T_C$ as the inflection point in the magnetization vs. temperature curve, the observed values are found to be 810 K, 890 K and 900 K for the LP573K, LP773K and LP673K samples, respectively. The $T_C$ value for the HP300K sample is similar to the value obtained for LP573. Using the theoretically calculated exchange interactions, $T_C$ for different degree of atomic order in CFA varying from B2 to L2$_1$ can be calculated using MC simulations. The volume was kept fixed as the degree of order varied between B2 and L2$_1$ and the data presented here represent the effects of differently ordered CFA phases. To obtain $T_C$ for the different phases, the occupancy of Fe atoms on the Heusler alloy 4a sites was varied from 50% to 100%, corresponding to changing the structure from B2 to L2$_1$. The estimated $T_C$ values, cf. Fig. 2 (b), monotonously increases from $T_C = 810$ K (B2) to $T_C = 950$ K (L2$_1$). A direct comparison between experimental and calculated $T_C$ values is hampered by the high temperature (beyond 800K) induced structural transition from well-ordered to partially-ordered CFA phase which interferes with the magnetic transition [39, 40]. The irreversible nature of the recorded magnetization vs. temperature curve indicates a distortion of structure for the ordered phase during measurement, even though interface alloying at elevated temperature cannot be ruled out. The experimentally observed $T_C$ values are presented in Table I.

### B. In-plane angle dependent FMR measurements

In-plane angle dependent FMR measurements were performed at 9.8 GHz frequency for all samples; the resonance field $H_r$ vs. in-plane angle $\phi_H$ of the applied magnetic field is plotted in Fig. 3. The experimental results have been fitted using the expression [41],

$$f = \frac{g_\parallel \mu_B \mu_0}{h}\left[\left\{H_r \cos(\phi_H - \phi_M) + \frac{2K_c}{\mu_0 M_s}\cos 4(\phi_M - \phi_c) + \frac{2K_u}{\mu_0 M_s}\cos 2(\phi_M - \phi_u)\right\}\left\{H_r \cos(\phi_H - \phi_M) + M_{eff} + \frac{K_c}{2\mu_0 M_s}(3 + \cos 4(\phi_M - \phi_c)) + \frac{2K_u}{\mu_0 M_s}\cos^2(\phi_M - \phi_u)\right\}\right]^{1/2}, \quad (1)$$

where $f$ is resonance frequency, $\mu_B$ is the Bohr magneton and $h$ is Planck constant. $\phi_M$, $\phi_u$ and $\phi_c$ are the in-plane directions of the magnetization, uniaxial anisotropy and cubic anisotropy, respectively, with respect to the [100] direction of the Si substrate. $H_u = \frac{2K_u}{\mu_0 M_s}$ and $H_c = \frac{2K_c}{\mu_0 M_s}$ are the in-plane uniaxial and cubic anisotropy fields, respectively, and $K_u$ and $K_c$ are the uniaxial and cubic magnetic anisotropy constants, respectively, $M_s$ is the saturation magnetization and $M_{eff}$ is the effective magnetization. By considering $\phi_H \sim \phi_M$, $H_u$ and $H_c$ $<<H_r<< M_{eff}$, equation (1) can be simplified as:

$$H_r = \left(\frac{hf}{\mu_0 g_\parallel \mu_B}\right)^2 \frac{1}{M_{eff}} - \frac{2K_c}{\mu_0 M_s}\cos 4(\phi_H - \phi_c) - \frac{2K_u}{\mu_0 M_s}\cos 2(\phi_H - \phi_u). \quad (2)$$

The extracted cubic anisotropy fields $\mu_0 H_c \leq 0.22$mT are negligible for all the samples. The extracted in-plane Landé splitting factors $g_\parallel$ and the uniaxial anisotropy fields $\mu_0 H_u$ are presented in Table I. The purpose of the angle dependent FMR measurements was only to investigate the symmetry of the in-plane magnetic anisotropy. Therefore, care was not taken to have the same in-plane orientation of the samples during angle dependent FMR measurements, which explains why the maxima appear at different angles for the different samples.

### C. Out-of-plane FMR measurements

Field-sweep out-of-plane FMR measurements were performed at different constant temperatures in the range 50K – 300K and at different constant frequencies in the range of 9-20 GHz. Figure 1(b) shows the amplitude of the complex transmission coefficient $S_{21}$(10 GHz) vs. field measured for the LP673K thin film at different temperatures. The recorded FMR spectra were fitted using the equation [42],

$$S_{21} = S\frac{(\Delta H/2)^2}{(H-H_r)^2+(\Delta H/2)^2} + A\frac{(\Delta H/2)(H-H_r)}{(H-H_r)^2+(\Delta H/2)^2} + D \cdot t, \quad (3)$$

where $S$ represents the coefficient describing the transmitted microwave power, $A$ is used to describe a waveguide induced phase shift contribution which is, however, minute, $H$ is applied magnetic field, $\Delta H$ is the full-width of half maximum, and $D \cdot t$ describes the linear drift in time ($t$) of the recorded signal. The extracted $\Delta H$ vs. frequency at different constant temperatures are shown in Fig. 4 for all the samples. For brevity only data at a few temperatures are plotted. The Gilbert damping was estimated using the equation [42],

$$\Delta H = \Delta H_0 + \frac{2h\alpha f}{g_\perp \mu_B \mu_0} \quad (4)$$

where $\Delta H_0$ is the inhomogeneous line-width broadening, $\alpha$ is the experimental Gilbert damping constant, and $g_\perp$ is the Landé splitting factor measured employing out-of-plane FMR. The insets in the figures show the temperature dependence of $\alpha$. The effective magnetization ($\mu_0 M_{eff}$) was estimated from the $f$ vs. $H_r$ curves using out-of-plane Kittel's equation [43],

$$f = \frac{g_\perp \mu_0 \mu_B}{h}(H_r - M_{eff}),  \quad (5)$$

as shown in Fig. 5. The temperature dependence of $\mu_0 M_{eff}$ and $\mu_0 \Delta H_0$ are shown as insets in each figure. The observed room temperature values of $\mu_0 M_{eff}$ are closely equal to the $\mu_0 M_s$ values obtained from static magnetization measurements, presented in Table I. The extracted values of $g_\perp$ at different temperatures are within error limits constant for all samples. However, the difference between estimated values of $g_\parallel$ and $g_\perp$ is ≤ 3%. This difference could stem from the limited frequency range used since these values are quite sensitive to the value of $M_{eff}$, and even a minute uncertainty in this quantity can result in the observed small difference between the $g_\parallel$ and $g_\perp$ values.

To obtain the intrinsic Gilbert damping ($\alpha_{int}$) all extrinsic contributions to the experimental $\alpha$ value need to be subtracted. In metallic ferromagnets, the intrinsic Gilbert damping is mostly caused by electron magnon scattering, but several other extrinsic contributions can also contribute to the experimental value of the damping constant. One contribution is two-magnon scattering which is however minimized for the perpendicular geometry used in this study and therefore this contribution is disregarded [44]. Another contribution is spin-pumping into the capping layer as the LP573K, LP673K and LP773K samples are capped with 4 nm of Al that naturally forms a thin top layer consisting of $Al_2O_3$. Since spin pumping in low spin-orbit coupling materials with thickness less than the spin-diffusion length is quite small this contribution is also disregarded in all samples. However, the HP300K sample is capped with Ta and therefore a spin-pumping contribution have been subtracted from the experimental $\alpha$ value; $\alpha_{sp} = \alpha_{HP300K}(\text{with Ta capping}) - \alpha_{HP300K}(\text{without capping}) \approx 1 \times 10^{-3}$. The third contribution arises from the inductive coupling between the precessing magnetization and the CPW, a reciprocal phenomenon of FMR, known as radiative damping $\alpha_{rad}$ [45]. This damping is directly proportional to the magnetization and thickness of the thin films samples and therefore usually dominates in thicker and/or high magnetization samples. The last contribution is eddy current damping ($\alpha_{eddy}$) caused by eddy currents in metallic ferromagnetic thin films [45, 46]. As per Faraday's law the time varying magnetic flux density generates an AC voltage in the metallic ferromagnetic layer and therefore results in the eddy current damping. This damping is directly proportional to the square of the film thickness and inversely proportional to the resistivity of the sample [45].

In contrast to eddy-current damping, $\alpha_{rad}$ is independent of the conductivity of the ferromagnetic layer, hence this damping mechanism is also operative in ferromagnetic insulators. Assuming a uniform magnetization of the sample the radiative damping can be expressed as [45],

$$\alpha_{rad} = \frac{\eta \gamma \mu_0^2 M_S \delta l}{2 Z_0 w}, \tag{6}$$

where $\gamma = g\mu_B/\hbar$ is the gyromagnetic ratio, $Z_0 = 50$ Ω is the waveguide impedance, $w = 240$ µm is the width of the waveguide, $\eta$ is a dimensionless parameter which accounts for the FMR mode profile and depends on boundary conditions, and $\delta$ and $l$ are the thickness and length of the sample on the waveguide, respectively. The strength of this inductive coupling depends on the inductance of the FMR mode which is determined by the waveguide width, sample length over waveguide, sample saturation magnetization and sample thickness. The dimensions of the LP573K, LP673K and LP773K samples were 6.3×6.3 mm², while the dimensions of the HP300K sample were 4×4 mm². The $\alpha_{rad}$ damping was estimated experimentally as explained by Schoen *et al.* [45] by placing a 200 µm thick glass spacer between the waveguide and the sample, which decreases the radiative damping by more than one order magnitude as shown in Fig. 6(a). The measured radiative damping by placing the spacer between the waveguide and the LP773 sample, $\alpha_{rad} = \alpha_{without\ spacer} - \alpha_{with\ spacer} \approx (2.36 \pm 0.10 \times 10^{-3}) - (1.57 \pm 0.20 \times 10^{-3}) = 0.79 \pm 0.22 \times 10^{-3}$. The estimated value matches well with the calculated value using Eq. (6); $\alpha_{rad} = 0.78 \times 10^{-3}$. Our results are also analogous to previously reported results on radiative damping [45]. The estimated temperature dependent radiative damping values for all samples are shown in Fig. 6(b).

Spin wave precession in ferromagnetic layers induces an AC current in the conducting ferromagnetic layer which results in eddy current damping. It can be expressed as [45, 46],

$$\alpha_{eddy} = \frac{C\gamma \mu_0^2 M_S \delta^2}{16\ \rho}, \tag{7}$$

where $\rho$ is the resistivity of the sample and $C$ accounts for the eddy current distribution in the sample; the smaller the value of $C$ the larger is the localization of eddy currents in the sample. The measured resistivity values between 300 K to 50 K temperature range fall in the ranges 1.175 – 1.145 µΩ-m, 1.055 – 1.034 µΩ-m, 1.035 – 1.00 µΩ-m, and 1.45 – 1.41 µΩ-m for the LP573K, LP673, LP773 and HP300K samples, respectively. The parameter $C$ was obtained from thickness dependent experimental Gilbert damping constants measured for B2 ordered films, by linear fitting of $\alpha - \alpha_{rad} \approx \alpha_{eddy}$ vs. $\delta^2$ keeping other parameters constant (cf. Fig. 6(c)). The fit to the data yielded $C \approx 0.5 \pm 0.1$. These results are concurrent to those obtained for permalloy thin films [45]. Since the variations of the resistivity and magnetization for the samples are small, we have used the same $C$ value for the estimation of the eddy current damping in all the samples. The estimated temperature dependent values of the eddy current damping are presented in Fig. 6(d).

All these contributions have been subtracted from the experimentally observed values of $\alpha$. The estimated intrinsic Gilbert damping $\alpha_{int}$ values so obtained are plotted in Fig. 7(a) for all samples.

### D. Theoretical results: first principle calculations

The calculated temperature dependent intrinsic Gilbert damping for $Co_2FeAl$ phases with different degree of atomic order are shown in Fig. 7(b). The temperature dependent Gilbert damping indicates that the lattice displacements and spin fluctuations contribute differently in the A2, B2 and $L2_1$ phases. The torque correlation model [47, 48] describes qualitatively two contributions to the Gilbert damping. The first one is the intraband scattering where the band index is always conserved. Since it has a linear dependence on the electron lifetime, in the low temperature regime this term increases rapidly, it is also known as the conductivity like scattering. The second mechanism is due to interband transitions where the scattering occurs between bands with different indices. Opposite to the intraband scattering, the resistivity like interband scattering with an inverse dependence on the electron lifetime increases with increasing temperature. The sum of the intra- and interband electron scattering contributions gives rise to a non-monotonic dependence of the Gilbert damping on temperature for the $L2_1$ structure. In contrast to the case for $L2_1$, only interband scattering is present in the A2 and B2 phases, which results in a monotonic increase of the intrinsic Gilbert damping with increasing temperature. This fact is also supported by a previous study [37] which showed that even a minute chemical disorder can inhibit the intraband scattering of the system. Our theoretical results manifest that the $L2_1$ phase has the lowest Gilbert damping around $4.6 \times 10^{-4}$ at 300 K, and that the value for the B2 phase is only slightly larger at room temperature. According to the torque correlation model, the two main contributions to damping are the spin orbit coupling and the density of states (DOS) at the Fermi level [47, 48]. Since the spin orbit strength is the same for the different phases it is enough to focus the discussion on the DOS that provides a qualitative explanation why damping is found lower in B2 and L2$_1$ structures compared to A2 structure. The DOS at the Fermi level of the B2 phase (24.1 states/Ry/f.u; f.u = formula unit) is only slightly larger to that of the L2$_1$ phase (20.2 states/Ry/f.u.), but both are significantly smaller than for the A2 phase (59.6 states/Ry/f.u.) as shown in Fig. 8. The gap in the minority spin channel of the DOS for the B2 and L2$_1$ phases indicate half-metallicity, while the A2 phase is metallic. The atomically resolved spin polarized DOS indicates that the Fermi-level states mostly have contributions from Co and Fe atoms. For transition elements such as Fe and Ni, it has been reported that the intrinsic Gilbert damping increases significantly below 100K with decreasing temperature [37]. The present electronic structure calculations were performed using Green's functions, which do rely on a phenomenological relaxation time parameter, on the expense that the different contributions to damping cannot be separated easily. The reported results in Ref. [37] are by some means similar to our findings of the temperature dependent Gilbert damping in full Heusler alloy films with different degree of atomic order. The intermediate states of B2 and L2$_1$ are more close to the trend of B2 than L2$_1$, which indicates that even a tiny atomic order induced by the Fe and Al site disorder will inhibit the conductivity-like channel in the low temperature region. The theoretically calculated Gilbert damping constants are matching qualitatively with the experimentally observed $\alpha_{int}$ values as shown in Fig. 7. However, the theoretically calculated $\alpha_{int}$ for the L2$_1$ phase increases rapidly below 100K, in contrast to the experimental results for the well-ordered CFA thin film (LP673K) indicating that $\alpha_{int}$ saturates at low temperature. This discrepancy between the theoretical and experimental results can be understood taking into account the low temperature behaviour of the life time $\tau$ of Bloch states. The present theoretical model assumed that the Gilbert damping has a linear dependence on the electron lifetime in intraband transitions which is however correct only in the limit of small lifetime, *i.e.*, $qv_F\tau \ll 1$, where $q$ is the magnon wave vector and $v_F$ is the electron Fermi velocity. However, in the low temperature limit the lifetime $\tau$ increases and as a result of the anomalous skin effect the intrinsic Gilbert damping saturates $\alpha_{int} \propto \tan^{-1}qv_F\tau / qv_F$ at low temperature [37], which is evident from our experimental results.

Remaining discrepancies between theoretical and experimental values of the intrinsic Gilbert damping might stem from the fact that the samples used in the present study are polycrystalline and because of sample imperfections these films exhibit significant inhomogenous line-width broadening due to superposition of local resonance fields.

**CONCLUSION**

In summary, we report temperature dependent FMR measurements on as-deposited $Co_2FeAl$ thin films with different degree of atomic order. The degree of atomic ordering is established by comparing experimental and theoretical results for the temperature dependent intrinsic Gilbert damping constant. It is evidenced that the experimentally observed intrinsic Gilbert damping in samples with atomic disorder (A2 and B2 phase samples) decreases with decreasing temperature. In contrast, the atomically well-ordered sample, which we identify at least partial $L2_1$ phase, exhibits an intrinsic Gilbert damping constant that increases with decreasing temperature. These temperature dependent results are explained employing the torque correction model including interband transitions and both interband as well as intraband transitions for samples with atomic disorder and atomically ordered phases, respectively.

**ACKNOWLEDGEMENT**

This work is supported by the Knut and Alice Wallenberg (KAW) Foundation, Grant No. KAW 2012.0031 and from Göran Gustafssons Foundation (GGS), Grant No. GGS1403A. The computations were performed on resources provided by SNIC (Swedish National Infrastructure for Computing) at NSC (National Supercomputer Centre) in Linköping, Sweden. S. H. acknowledges the Department of Science and Technology India for providing the INSPIRE fellow (IF140093) grant. Daniel Hedlund is acknowledged for performing magnetization versus temperature measurements.

**Author Information**

Corresponding Authors E-mails: ankit.kumar@angstrom.uu.se, peter.svedlindh@angstrom.uu.se

**REFERENECS**

1. S. Picozzi, A. Continenza, and A. J. Freeman, Phys. Rev. B. **69**, 094423 (2004).
2. I. Galanakis, P. H. Dederichs, and N. Papanikolaou, Phys. Rev. B **66**, 174429 (2002).
3. Z. Bai, L. Shen, G. Han, Y. P. Feng, Spin **02**, 1230006 (2012).
4. M. Belmeguenai, H. Tuzcuoglu, M. S. Gabor, T. Petrisor jr, C. Tuisan, F. Zighem, S. M. Chérif, P. Moch, J. Appl. Phys. **115**, 043918 (2014).

5. I. Galanakis, P. H. Dederichs, and N. Papanikolaou, Phys. Rev. B **66**, 174429 (2002).
6. S. Skaftouros, K. Ozdogan, E. Sasioglu, I. Galanakis, Physical Review B **87**, 024420 (2013).
7. J. Chico, S. Keshavarz, Y. Kvashnin, M. Pereiro, I. D. Marco, C. Etz, O. Eriksson, A. Bergman, and L. Bergqvist, Phys. Rev. B **93**, 214439 (2016).
8. B. Pradines, R. Arras, I. Abdallah, N. Biziere, and L. Calmels, Phys. Rev. B **95**, 094425 (2017).
9. X. G. Xu, D. L. Zhang, X. Q. Li, J. Bao, Y. Jiang, , and M. B. A. Jalil, J. Appl. Phys. **106**, 123902 (2009).
10. M. Belmeguenai, H. Tuzcuoglu, M. S. Gabor, T. Petrisor, Jr., C. Tiusan, D. Berling, F. Zighem, T. Chauveau, S. M. Chérif, and P. Moch, Phys. Rev. B **87**, 184431 (2013).
11. S. Qiao, S. Nie, J. Zhao, Y. Huo, Y. Wu, and X. Zhang, Appl. Phys. Lett. **103**, 152402 (2013).
12. B. S. Chun, K. H. Kim, N. Leibing, S. G. Santiago, H. W. Schumacher, M. Abid, I. C. Chu, O. N. Mryasov, D. K. Kim, H. C. Wu, C. Hwang, and Y. K. Kim, Acta Mater. **60**, 6714 (2012).
13. H. Sukegawa, Z. C. Wen, K. Kondou, S. Kasai, S. Mitani, and K. Inomata, Appl. Phys. Lett. **100**, 182403 (2012).
14. G. Aston, H. R. Kaufman, and P. J. Wilbur, AIAAA journal **16**, 516–524 (1978).
15. A. Kumar, D. K. Pandya, and S. Chaudhary, J. Appl. Phys. **111**, 073901 (2012).
16. S. Husain, S. Akansel, A. Kumar, P. Svedlindh, and S. Chaudhary, Scientific Reports **6**, 28692 (2016).
17. A Sakuma, J. Phys. D: Appl. Phys. **48**, 164011 (2015).
18. S. Husain, A. Kumar, S. Chaudhary, and P. Svedlindh, AIP Conference Proceedings 1728, 020072 (2016).
19. Y. Takamura, R. Nakane, and S. Sugahare, J. Appl. Phys. **105**, 07B109 (2009).
20. S. Husain, A. Kumar, S. Akansel, P. Svedlindh, and S. Chaudhary, J. Mag. Mag. Mater. **442**, 288–294 (2017).
21. K. Inomata, S. Okamura, A. Miyazaki1, M. Kikuchi, N. Tezuka, M. Wojcik and E. Jedryka, J. Phys. D: Appl. Phys. **39**, 816–823 (2006).
22. M. Oogane, Y. Sakuraba, J. Nakata1, H. Kubota, Y. Ando, A. Sakuma, and T. Miyazaki, J. Phys. D: Appl. Phys. **39**, 834-841 (2006).
23. S. Okamura, A. Miyazaki, N. Tezuka, S. Sugimoto, and K. Inomata, Materials Transactions **47**, 15 - 19 (2006).
24. H. Ebert, D. Ködderitzsch, and J. Minár, Rep. Prog. Phys. 74, 096501 (2011).
25. H. Ebert, http://ebert.cup.uni-muenchen.de/SPRKKR.
26. S. H. Vosko, L. Wilk, and M. Nusair, Canadian J. Phys. **58**, 1200 (1980).
27. M. Belmeguenai, H. Tuzcuoglu, M. Gabor, T. Petrisor, C. Tiusan, D. Berling, F. Zighem, and S. M. Chrif, J. Mag. Mag. Mat. **373**, 140 (2015).
28. A. Liechtenstein, M. Katsnelson, V. Antropov, and V. Gubanov, J. Mag. Mag. Mat. **67**, 65 (1987).
29. H. Ebert, and S. Mankovsky, Phys. Rev. B **79**, 045209 (2009).

30. N. Metropolis, A. W. Rosenbluth, M. N. Rosenbluth, A. H. Teller, and E. Teller, J. Chem. Phys. **21**, 1087 (1953).
31. B. Skubic, J. Hellsvik, L. Nordström, and O. Eriksson, J. Phys.: Cond. Matt. **20**, 315203 (2008).
32. http://physics.uu.se/uppasd.
33. P. Soven, Phys. Rev. **156**, 809 (1967).
34. G. M. Stocks, W. M. Temmerman, and B. L. Gyorffy, Phys. Rev. **41**, 339 (1978).
35. A. Brataas, Y. Tserkovnyak, and G. E. W. Bauer, Phys. Rev. Lett. **101**, 037207 (2008).
36. H. Ebert, S. Mankovsky, D. Ködderitzsch, and P. J. Kelly, Phys. Rev. Lett. **107**, 066603 (2011).
37. S. Mankovsky, D. Ködderitzsch, G. Woltersdorf, and H. Ebert, Phys. Rev. B **87**, 014430 (2013).
38. H. Ebert, S. Mankovsky, K. Chadova, S. Polesya, J. Minár, and D. Ködderitzsch, Phys. Rev. B **91**, 165132 (2015).
39. Rie Y. Umetsu, A. Okubo, M. Nagasako, M. Ohtsuka, R. Kainuma, and K. Ishida, Spin **04**, 1440018 (2014).
40. D. Comtesse, B. Geisler, P. Entel, P. Kratzer, and L. Szunyogh, Phys. Rev. B **89**, 094410 (2014).
41. H. Kurebayashi, T. D. Skinner, K. Khazen, K. Olejník, D. Fang, C. Ciccarelli, R. P. Campion, B. L. Gallagher, L. Fleet, A. Hirohata, and A. J. Ferguson, Appl. Phys. Lett. **102**, 062415 (2013).
42. Y. Zhao, Qi Song, S.-H. Yang, T. Su, W. Yuan, S. S. P. Parkin, J Shi & W. Han, Scientific Reports **6**, 22890 (2016).
43. J. M. Shaw, Hans T. Nembach, T. J. Silva, C. T. Boone, J. Appl. Phys. **114**, 243906 (2013).
44. P. Landeros, R. E. Arias, and D. L. Mills, Phys. Rev. B 77, 214405 (2008).
45. M. A. W. Schoen, J. M. Shaw, H. T. Nembach, M. Weiler, and T. J. Silva, Phys. Rev. B **92**, 184417 (2015).
46. Y. Li and W. E. Bailey, Phys. Rev. Lett. **116**, 117602 (2016).
47. V. Kamberský, Czech. J. Phys. B **26**, 1366 (1976).
48. K. Gilmore, Y. U. Idzerda, and M. D. Stiles, Phys. Rev. Lett. **99**, 027204 (2007).

Table I Parameters describing magnetic properties of the different CFA samples.

| Sample | $\mu_0 M_S (\mu_0 M_{eff})$ (T) | $\mu_0 H_{ci}$ (mT) | $g_\parallel (g_\perp)$ | $\mu_0 H_u$ (mT) | $T_C$ (K) | $\alpha_{int}$ ($\times 10^{-3}$) |
|---|---|---|---|---|---|---|
| LP573K | 1.2±0.1 (1.091±0.003) | 0.75 | 2.06 (2.0) | 1.56 | 810 | 2.56 |
| LP673K | 1.2±0.1 (1.110±0.002) | 0.57 | 2.05 (2.0) | 1.97 | >900 | 0.76 |
| LP773K | 1.2±0.1 (1.081±0.002) | 0.46 | 2.05 (2.0) | 1.78 | 890 | 1.46 |
| HP300K | 0.9±0.1 (1.066±0.002) | 1.32 | 2.01 (2.0) | 3.12 | -- | 3.22 |

**Figure 1**

**Fig. 1.** (a) Layout of the in-house made VNA-based out-of-plane ferromagnetic resonance setup. (b) Out-plane ferromagnetic resonance spectra recorded for the well-ordered LP673K sample at different temperatures $f = 10$ GHz.

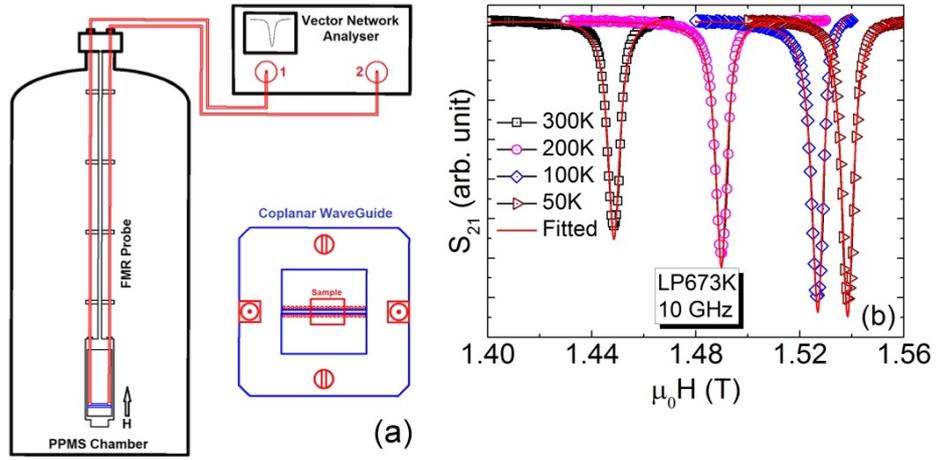

**Figure 2**

**Fig. 2.** (a) Magnetization vs. temperature plots measured on the CFA films with different degree of atomic order. (b) Theoretically calculated magnetization vs. temperature curves for CFA phases with different degree of atomic order, where 50 % (100 %) Fe atoms on Heusler alloy 4a sites indicate B2 (L2$_1$) ordered phase, and the rest are intermediate B2 & L2$_1$ mixed ordered phases.

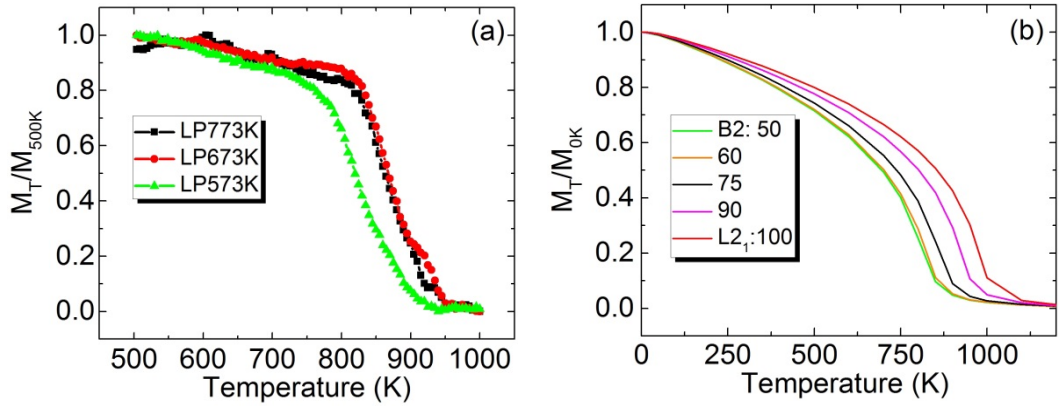

**Figure 3**

**Fig. 3.** Resonance field vs. in-plane orientation of the applied magnetic field of (a) $T_S = 300°C$, $P_{Ion} = 75\,W$ deposited, (b) $T_S = 400°C$, $P_{Ion} = 75\,W$ deposited, (c) $T_S = 500°C$, $P_{Ion} = 75\,W$ deposited, and (d) $T_S = 27°C$, $P_{Ion} = 100\,W$ deposited films. Red lines correspond to fits to the data using Eq. (1).

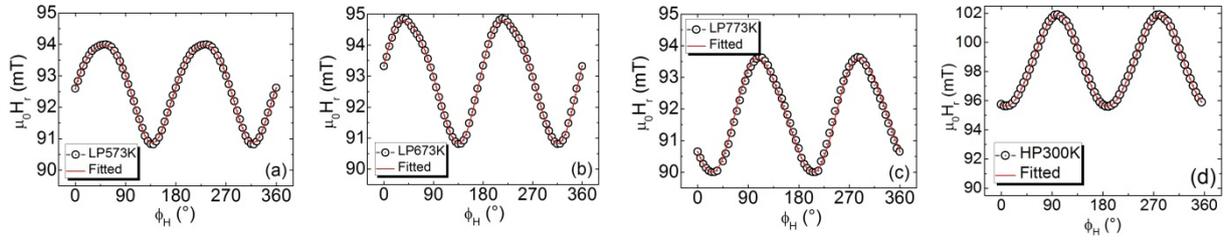

**Fig. 4.** Line-width vs. frequency of (a) $T_S = 300°C$, $P_{Ion} = 75\,W$ deposited, (b) $T_S = 400°C$, $P_{Ion} = 75\,W$ deposited, (c) $T_S = 500°C$, $P_{Ion} = 75\,W$ deposited, and (d) $T_S = 27°C$, $P_{Ion} = 100\,W$ deposited samples. Red lines correspond to fits to the data to extract the experimental Gilbert damping constant and inhomogeneous line-width. Respective insets show the experimentally determined temperature dependent Gilbert damping constants.

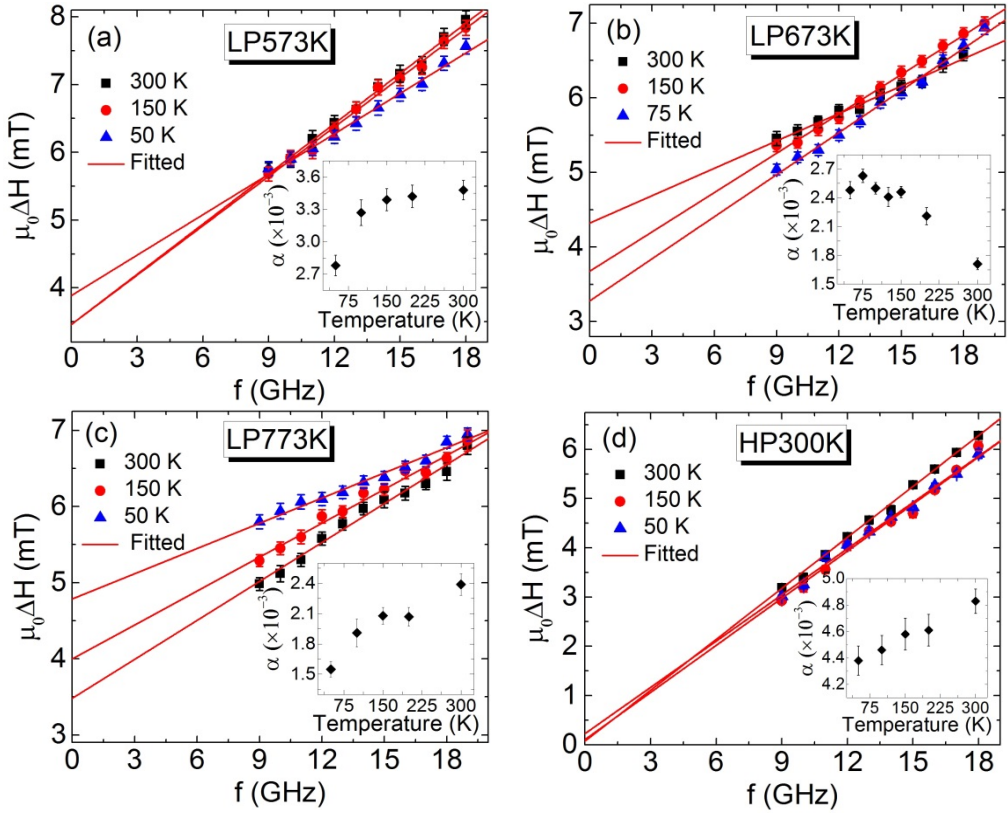

**Figure 5**

**Fig. 5.** Frequency vs. applied field of (a) $T_S = 300°C$, $P_{Ion} = 75\,W$ deposited, (b) $T_S = 400°C$, $P_{Ion} = 75\,W$ deposited, (c) $T_S = 500°C$, $P_{Ion} = 75\,W$ deposited, and (d) $T_S = 27°C$, $P_{Ion} = 100\,W$ deposited samples. Red lines correspond to Kittel's fits to the data. Respective insets show the temperature dependent effective magnetization and inhomogeneous line-width broadening values.

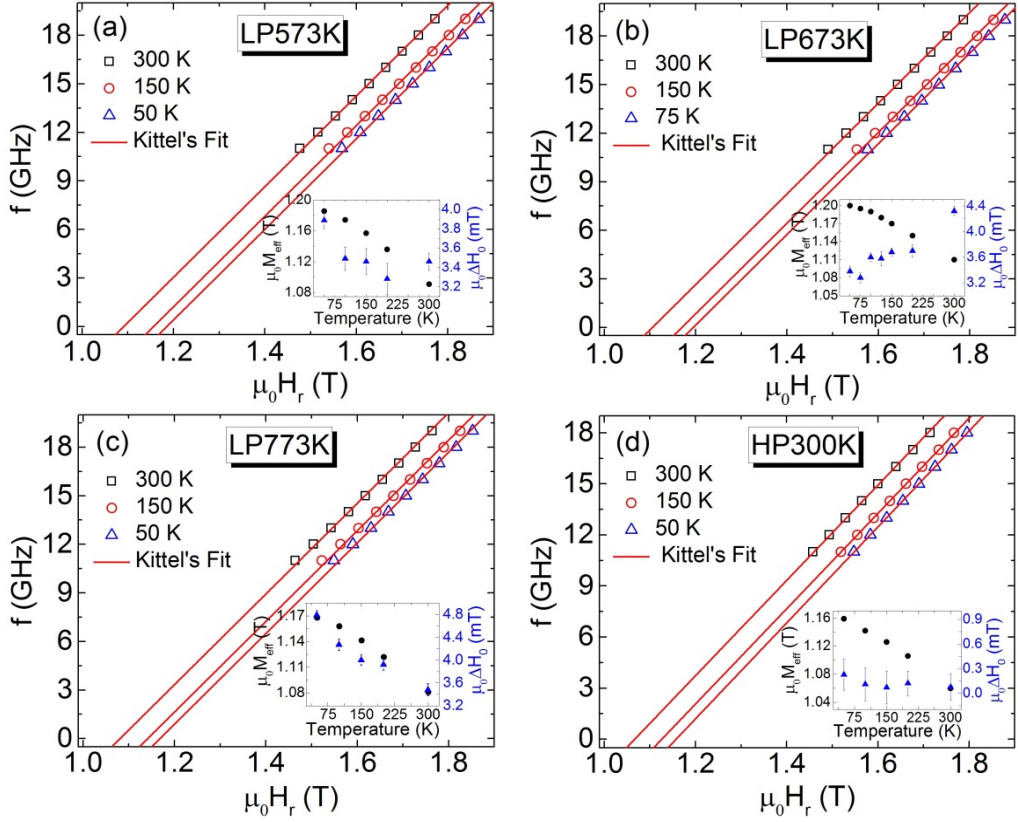

**Figure 6**

**Fig. 6.** (a) Linewidth vs. frequency with and without a glass spacer between the waveguide and the sample. Red lines correspond to fits using Eq. (4). (b) Temperature dependent values of the radiative damping using Eq. (6). The lines are guide to the eye. (c) $\alpha - \alpha_{rad} \approx \alpha_{eddy}$ vs $\delta^2$. The red line corresponds to a fit using Eq. (7) to extract the value of the correction factor $C$. (d) Temperature dependent values of eddy current damping using Eq. (7). The lines are guide to the eye.

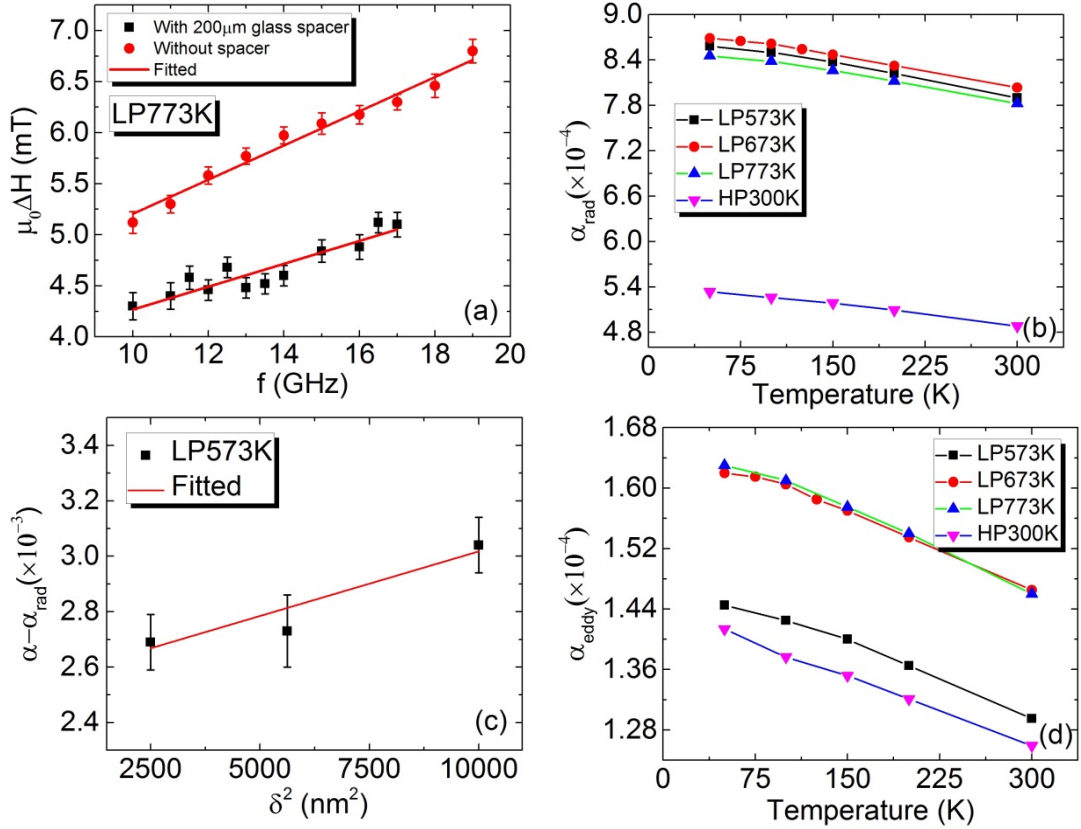

**Figure 7**

**Fig. 7.** Experimental (a) and theoretical (b) results for the temperature dependent intrinsic Gilbert damping constant for CFA samples with different degree of atomic order. The B2 & L2$_1$ mixed phase corresponds to the 75 % occupancy of Fe atoms on the Heusler alloy 4a sites. The lines are guide to the eye.

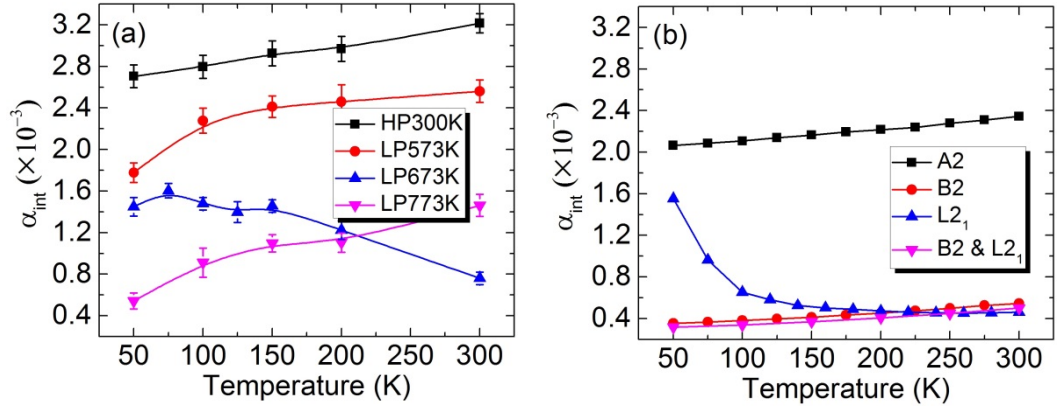

**Figure 8**

**Fig. 8.** Total and atom-resolved spin polarized density of states plots for various compositional CFA phases; (a) A2, (b) B2 and (c) L2$_1$.

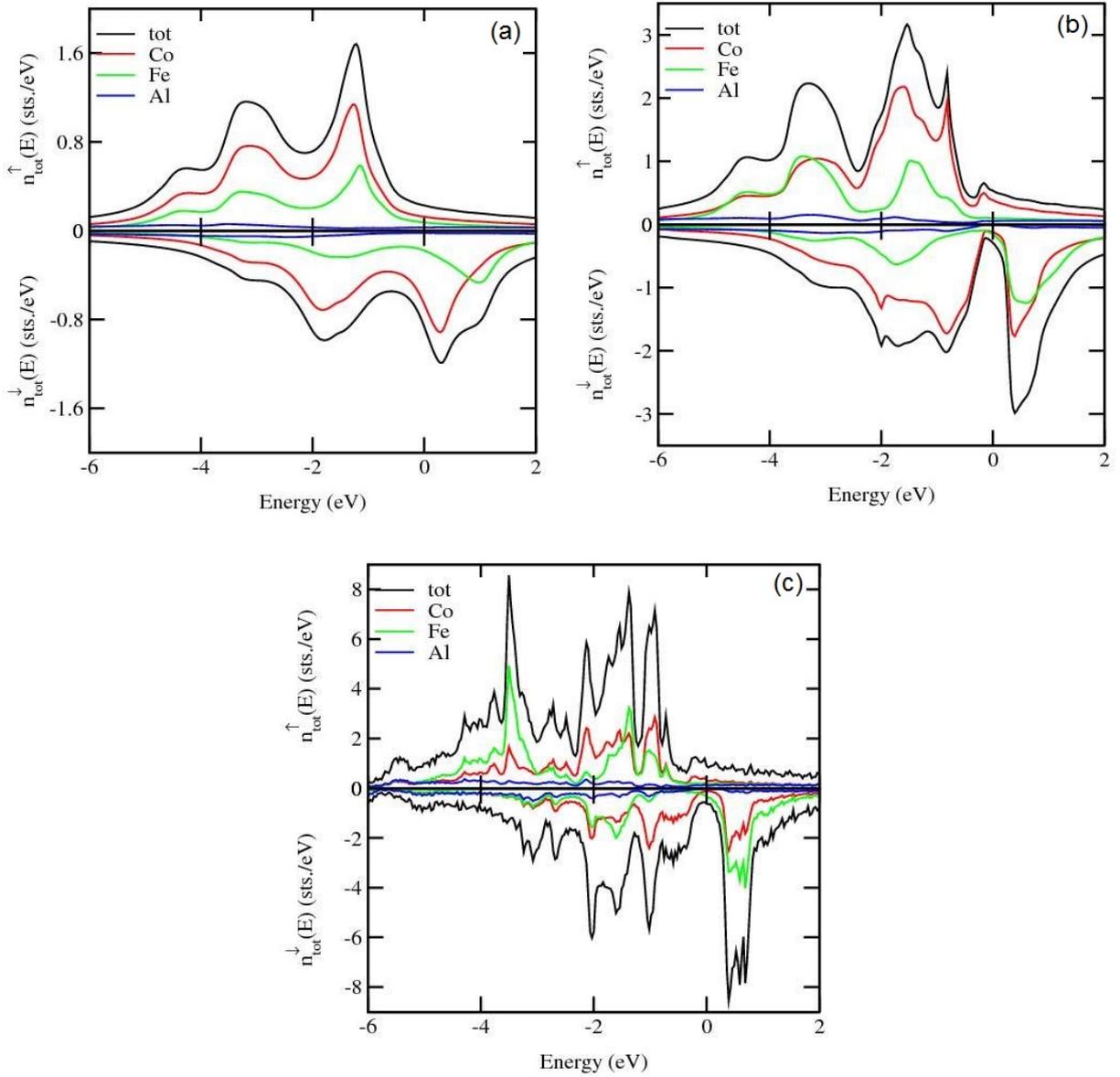